\newif\ifuseneurips
\useneuripsfalse  

\ifuseneurips
\PassOptionsToPackage{authoryear, sort&compress, round}{natbib}
\documentclass{article}
\usepackage[position]{neurips_2026}
\else
\documentclass[11pt, a4paper, logo]{googlecloud}
\fi

\pdftrailerid{redacted}

\ifuseneurips
\else
\makeatletter
\renewcommand\bibentry[1]{\nocitep{#1}{\frenchspacing\@nameuse{BR@r@#1\@extra@b@citeb}}}
\makeatother
\fi

\usepackage{amsmath,amssymb,amsfonts}
\usepackage{kantlipsum, lipsum}
\usepackage{dsfont}
\ifuseneurips
\else
\usepackage[authoryear, sort&compress, round]{natbib}
\fi

\usepackage[utf8]{inputenc} %
\usepackage[T1]{fontenc}    %
\usepackage{hyperref}       %
\hypersetup{
  colorlinks,
  citecolor=blue,
  linkcolor=red,
  urlcolor=blue}
\usepackage{booktabs}       %
\usepackage{nicefrac}       %
\usepackage{microtype}      %
\usepackage{wrapfig} %
\usepackage{soul} %
\usepackage{tikz, adjustbox}
\usepackage{arydshln}
\usepackage[capitalize,noabbrev]{cleveref}
\usepackage{fancyvrb}

\usepackage{inconsolata}
\usepackage[skins,breakable,most]{tcolorbox}

\usepackage{caption}
\usepackage{subcaption}
\usepackage{enumitem}
\usepackage{flushend}
\usepackage{balance}
\usepackage{lineno}
\usepackage{amsmath,amssymb,amsfonts}
\usepackage{algorithm}
\usepackage{algpseudocode}
\usepackage{graphicx}
\usepackage{textcomp}
\usepackage{xcolor}
\usepackage{xspace}
\usepackage{colortbl}
\usepackage{amsthm}
\usepackage{url}
\usepackage{float}
\usepackage{multirow}
\usepackage{multicol}
\usepackage{color}
\usepackage{bm}
\usepackage{bbm}

\usepackage{pifont}
\newcommand{\cmark}{\ding{51}}%
\newcommand{\xmark}{\ding{55}}%
\newcommand{\pmark}{$\boldsymbol{\sim}$}%

\usepackage{array}
\newcolumntype{L}[1]{>{\raggedright\let\newline\\\arraybackslash\hspace{0pt}}m{#1}}
\newcolumntype{C}[1]{>{\centering\let\newline  \\\arraybackslash\hspace{0pt}}m{#1}}
\newcolumntype{R}[1]{>{\raggedleft\let\newline \\\arraybackslash\hspace{0pt}}m{#1}}

\definecolor{lightorange}{RGB}{245, 237, 211}
\definecolor{clovergreen}{RGB}{32,115,55}

\newtcbox{\hlprimarytab}{on line, rounded corners, box align=base, colback=c3!10,colframe=white,size=fbox,arc=3pt, before upper=\strut, top=-2pt, bottom=-4pt, left=-2pt, right=-2pt, boxrule=0pt}
\newtcbox{\hlsecondarytab}{on line, box align=base, colback=blue!10,colframe=white,size=fbox,arc=3pt, before upper=\strut, top=-2pt, bottom=-4pt, left=-2pt, right=-2pt, boxrule=0pt}
\newtcbox{\hlcasetab}{on line, box align=base, colback=c5!10,colframe=white,size=fbox,arc=3pt, before upper=\strut, top=-2pt, bottom=-4pt, left=-2pt, right=-2pt, boxrule=0pt}

\definecolor{c1}{cmyk}{0,0.6175,0.8848,0.1490}
\definecolor{c2}{cmyk}{0.1127,0.6690,0,0.4431}
\definecolor{c3}{cmyk}{0.3081,0,0.7209,0.3255}
\definecolor{c4}{cmyk}{0.6765,0.2017,0,0.0667}
\definecolor{c5}{cmyk}{0,0.8765,0.7099,0.3647}

\DeclareMathOperator*{\argmax}{argmax}

\usepackage[utf8]{inputenc} 
\usepackage[T1]{fontenc}    
\usepackage{amsfonts}       
\usepackage{comment}
\usepackage{mdframed}
\usepackage{fontawesome}
\usepackage{csquotes}
\usepackage{makecell}
\definecolor{beigecolor}{RGB}{253, 244, 204} 
\definecolor{greencolor}{RGB}{228, 242, 217} 
\definecolor{bluecolor}{RGB}{66, 133, 244}
\definecolor{orgcolor}{RGB}{255, 140, 15}

\definecolor{redcolor}{RGB}{234, 67, 53}

\definecolor{ggreen}{RGB}{52, 168, 83}

\definecolor{gyellow}{RGB}{251, 188, 5}

\definecolor{lightorange}{RGB}{245, 237, 211}
\definecolor{bluebar}{RGB}{138,159,201}
\definecolor{pinkbar}{RGB}{232,180,189}

\usepackage{mathtools}

\usepackage{parskip}
\usepackage{titletoc}

\newcommand{\stitle}[1]{\noindent \textbf{#1.}}

\lstdefinestyle{mystyle}{
    backgroundcolor=\color{backcolour},
    commentstyle=\color{codegreen},
    keywordstyle=\color{magenta},
    numberstyle=\tiny\color{codegray},
    stringstyle=\color{codepurple},
    basicstyle=\ttfamily\scriptsize,
    breakatwhitespace=false,
    breaklines=true,
    captionpos=b,
    keepspaces=true,
    numbers=left,
    numbersep=5pt,
    showspaces=false,
    showstringspaces=false,
    showtabs=false,
    tabsize=2,
    frame=none,
    aboveskip=1pt,
    belowskip=1pt,
}
\lstdefinestyle{plainins}{
    backgroundcolor=\color{white},
    commentstyle=\color{codegreen},
    keywordstyle=\color{magenta},
    numberstyle=\tiny\color{codegray},
    stringstyle=\color{codepurple},
    basicstyle=\ttfamily\scriptsize,
    breakatwhitespace=false,
    breaklines=true,
    captionpos=b,
    keepspaces=true,
    numbers=none,
    numbersep=5pt,
    showspaces=false,
    showstringspaces=false,
    showtabs=false,
    tabsize=2,
    aboveskip=0pt,
    belowskip=0pt,
    frame=single
}
\lstdefinestyle{plainexam}{
    backgroundcolor=\color[HTML]{FFFCF3},
    commentstyle=\color{codegreen},
    keywordstyle=\color{magenta},
    numberstyle=\tiny\color{codegray},
    stringstyle=\color{codepurple},
    basicstyle=\ttfamily\scriptsize,
    breakatwhitespace=false,
    breaklines=true,
    captionpos=b,
    keepspaces=true,
    numbers=none,
    numbersep=5pt,
    showspaces=false,
    showstringspaces=false,
    showtabs=false,
    tabsize=2,
    aboveskip=0pt,
    belowskip=0pt
}

\title{Agentic Coding Needs Proactivity, Not Just Autonomy}

\ifuseneurips
\author{}
\else
\correspondingauthor{}
\renewcommand{\today}{}

\author{Nghi D. Q. Bui}
\author{Georgios Evangelopoulos}
\affil{Google Labs}
\begin{abstract}
Coding agents are rapidly changing the landscape of software development, moving from inline completion to autonomous systems that edit repositories, open pull requests, respond to issues, and run scheduled or webhook triggered routines across the development life cycle. The next generation is increasingly described as proactive and long-horizon: agents should notice relevant changes before the developer asks, connect signals across tools, decide when to interrupt, and carry preferences across sessions. Yet the field still lacks a clear account of what proactivity means for software development, how it differs from autonomy, what acceptance criteria proactive long-horizon tasks should satisfy, and which metrics determine whether unsolicited agent behavior is useful rather than merely active. Proactive coding agents should be evaluated by the quality and improvement of their \emph{insight policy}: the policy that decides what matters next, what evidence supports it, whether to show it, and how to adapt after feedback. This view is grounded in the principles of mixed initiative interaction. We propose a three level taxonomy of proactivity (\textbf{Reactive, Scheduled, and Situation Aware}), compare contemporary coding agents against five practical criteria, and sketch an active user simulation protocol with three evaluation targets: \textbf{Insight Decision Quality (IDQ)}, \textbf{Context Grounding Score (CGS)}, and \textbf{Learning Lift (LL)}.

\end{abstract}
\fi

\begin{document}

\maketitle

\ifuseneurips
\else
\makeatletter
{\renewcommand\thefootnote{}%
 \footnotetext{\hspace*{-1.8em}Correspondence: \texttt{\{nghib, gevang\}@google.com}}}
\makeatother
\fi

\section{Introduction}
\label{sec: intro}

Coding assistants have moved from inline completion to autonomous coding agents that participate in the full software development life cycle (SDLC)~\citep{tufano2024autodev,jin2024llmagentsse,li2025aidev,sase_survey}. The next shift is toward proactive agents that continuously absorb repository, toolchain, and workflow context, infer what matters from high-level developer needs, identify emerging problems or opportunities, and decide what to do next before a narrowly specified prompt arrives. This trajectory did not happen in one step: language model agents learned to combine reasoning with external actions~\citep{yao2023react} and executable code changes~\citep{wang2024codeact}; software engineering gave those agents repository interfaces, real issue-resolution benchmarks, and reusable platforms~\citep{yang2024sweagent,jimenez2024swebench,wang2024openhands}; and deployed tools moved them into the places where developers already work. Claude Code~\citep{anthropic2025claudecode}, OpenAI Codex~\citep{openai2025codex}, Gemini CLI~\citep{google2025geminicli}, Jules~\citep{google_jules2025_blog}, Google Antigravity~\citep{google_antigravity2025}, and OpenDev~\citep{bui2026building} place agents in terminals, IDEs, repositories, pull requests, and remote virtual machines. The resulting systems can edit files, run commands, inspect failures, and produce patches; the open question is no longer whether an agent can act, but whether it can decide when action should start and which contribution matters next.

Recent products also move initiation away from the explicit prompt. Cursor Automations~\citep{cursor_automations2026,techcrunch_cursor2026}, Claude Code Routines~\citep{anthropic_routines2026}, and Jules Scheduled Tasks~\citep{google_jules2025,jules_changelog} let coding agents run from schedules, webhooks, GitHub events, integrations, or monitored repository state. These are important triggers, but not yet situation aware proactivity. Triggered runs still differ from agents that notice context shifts on their own, infer whether those shifts matter, and choose whether to notify, question, draft, or stay silent. One stream of work treats proactivity as anticipating useful tasks from user activity and environment state~\citep{luProactiveAgentShifting2024,yangContextAgentContextAwareProactive2025,yangProAgentHarnessingOnDemand2025}; another treats it as asking before acting on uncertain intent~\citep{gan2024clarqllm,sun2025trainingproactivepersonalizedllm,zhou2026tomswe}. Both fit the mixed initiative framework of \citet{horvitz1999mixedinitiative}: reason about interruption cost, expected utility, and the benefit of leaving the user in control. Public documentation shows substantial autonomy and trigger coverage, but no clear evidence that production coding agents compute interruption cost, treat silence as an explicit action, or update what they show after developer feedback.

Proactive coding agents should be evaluated by the quality and improvement of their \emph{insight policy}, not by autonomous task completion alone. We call the unit of proactive behavior an \textbf{\emph{insight}}: a context grounded, time sensitive hypothesis about what matters next for the developer, paired with a decision to \emph{notify}, \emph{question}, \emph{draft}, or \emph{stay silent}. This makes the central question sharper. A proactive coding agent is not merely an agent that can do more work. It is an agent that learns which potential contributions deserve the developer's attention, which evidence makes them credible, and which future decisions should change after feedback.

\begin{figure}[t]
\centering
\includegraphics[width=\linewidth]{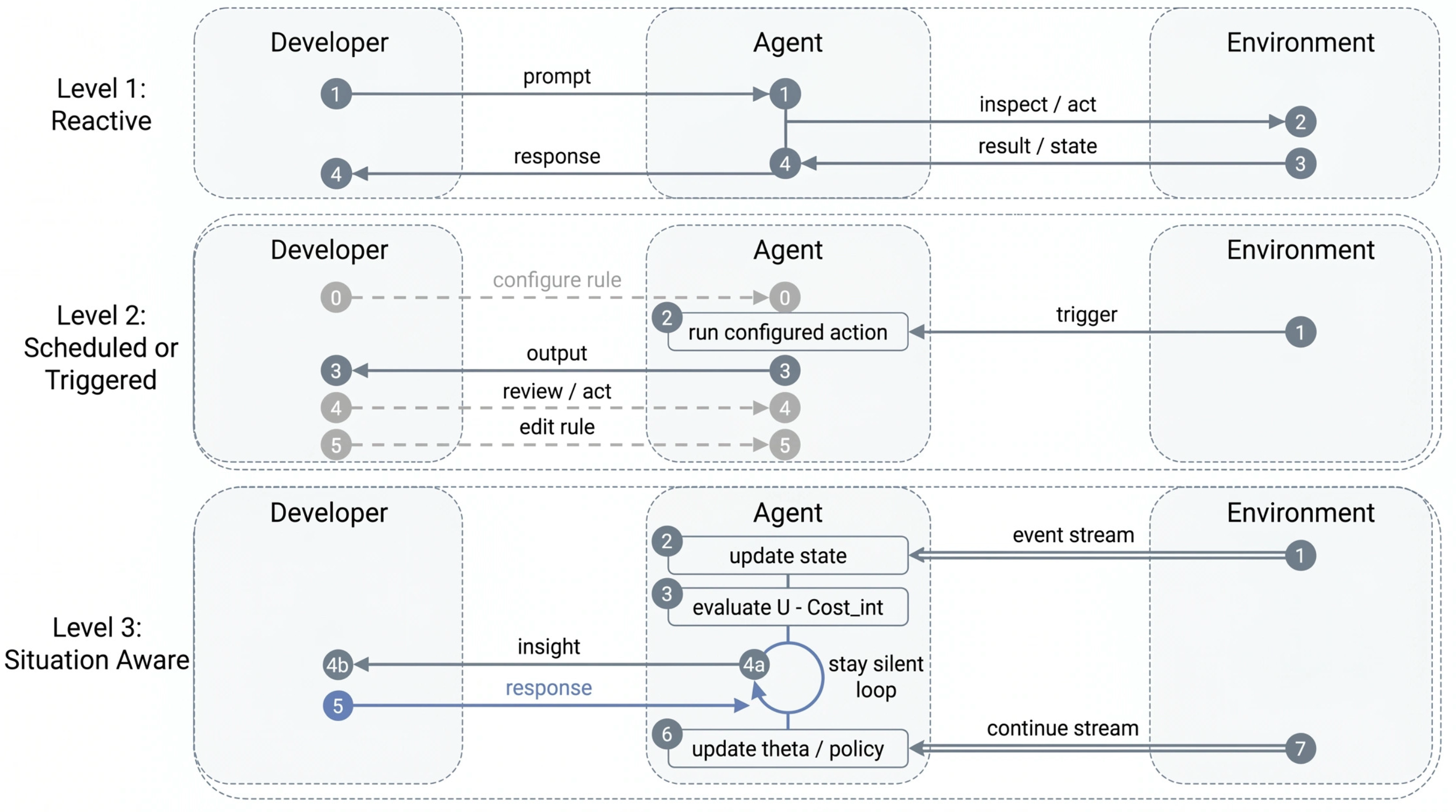}
\caption{Three levels of proactivity as numbered interaction flows. Level 1 begins with a developer prompt and ends after the agent response. Level 2 begins with a predefined schedule or event trigger; the developer may respond to the produced artifact, but this response does not automatically update a learned interruption policy. Level 3 monitors context continuously, chooses between staying silent and showing an insight, and uses developer feedback to update future timing, action selection, and framing.}
\label{fig:levels}
\end{figure}

To make this gap precise we propose a three level taxonomy, drawn from \citet{horvitz1999mixedinitiative} and \citet{handler2023balancing}. \textbf{Level 1, Reactive} agents run only when prompted. \textbf{Level 2, Scheduled} agents run from schedules or predefined triggers and may filter, batch, or rank outputs, but do not learn a cross context, per developer interruption policy. \textbf{Level 3, Situation Aware} agents monitor a continuous event stream, compare expected benefit with interruption cost, treat silence as an explicit action, and update a per developer model from feedback. Figure~\ref{fig:levels} expresses the taxonomy as numbered interaction flows to separate three properties that are easy to conflate: what initiates the agent, whether the developer may respond, and whether that response changes future policy. This distinction matters most for Level 2: scheduled automation may still produce artifacts that developers review, accept, dismiss, or edit, but that response is not equivalent to a learned policy over when to interrupt, ask, draft, or remain silent. The systems audited in Section~\ref{sec: experiments} cluster around Level 2; the gap is judgment: when to interrupt, what to show, and when to remain silent.

\begin{figure}[t]
\centering
\includegraphics[width=\linewidth]{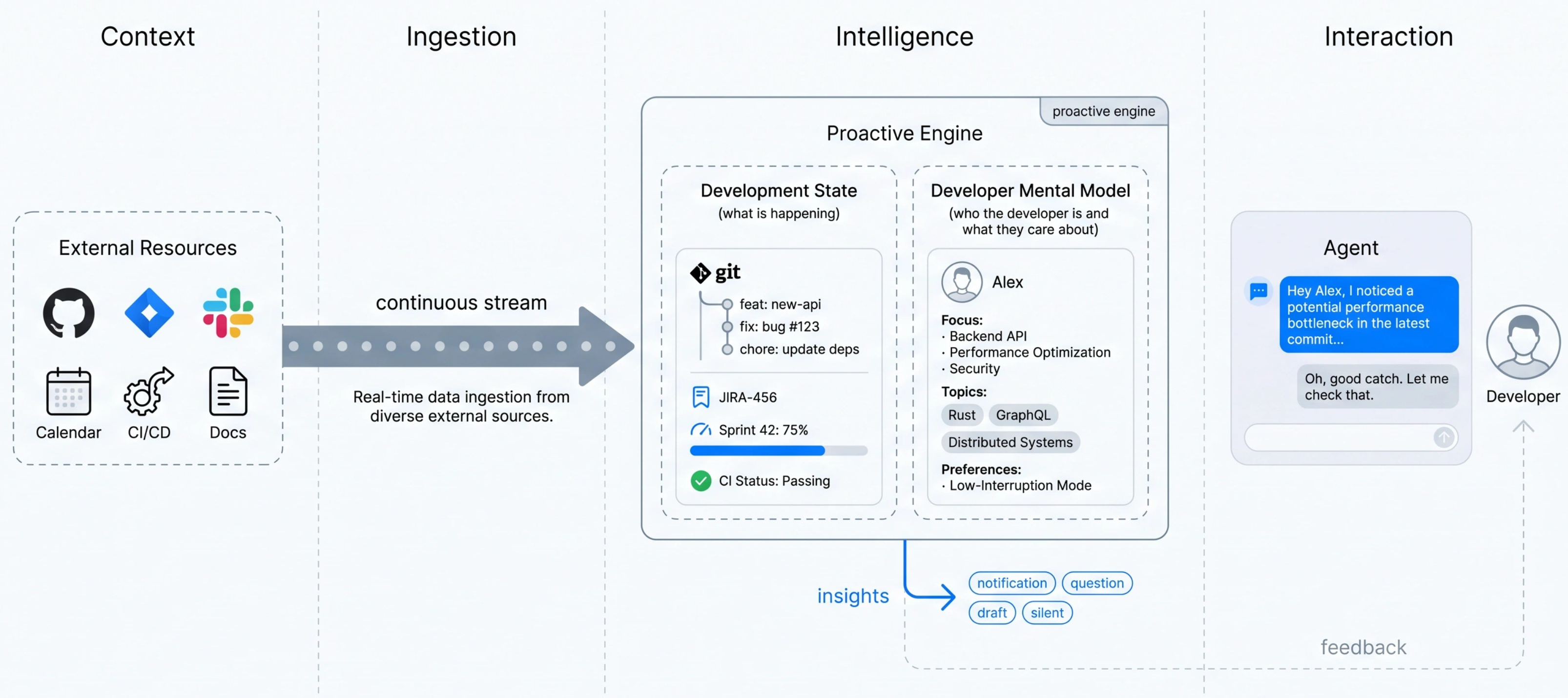}
\caption{Prototype proactive agentic coding engine. Context streams into an engine that maintains development state and a developer model, \textbf{emits \emph{insights}} (notify, question, draft, stay silent), and learns from response.}
\label{fig:overview}
\end{figure}

Figure~\ref{fig:overview} sketches a Level 3 engine: context streams into an engine that maintains development state and a developer mental model, \textbf{emits \emph{insights}} in the four actions used later (notify, question, draft, stay silent), and learns from response. A Level 3 engine is always deciding, but not always speaking; the quality of its insight policy is what a proactivity benchmark should measure.

The contribution is a framework for making proactive coding agents easier to define, compare, and evaluate. Rather than treating proactivity as a loose product label, the paper turns it into a set of design commitments and measurement targets centered on insights, interruption cost, silence, and learning from developer feedback. Concretely, the paper contributes:
\begin{itemize}[leftmargin=*,nosep]
\item A three level taxonomy separating autonomy from proactivity in agentic coding (Section~\ref{sec: method}).
\item A comparison of five contemporary coding agents against five practical criteria (Section~\ref{sec: experiments}).
\item A prototype Level 3 engine organized around \emph{insights} and the policy that selects, grounds, and updates them (Section~\ref{sec: method}).
\item Three insight-stream metrics adapted from active user simulation~\citep{nathani2026proactive}: Insight Decision Quality (IDQ), Context Grounding Score (CGS), and Learning Lift (LL) (Section~\ref{sec: experiments}).
\end{itemize}

\section{Background and Related Work}
\label{sec: related_work}
\label{sec: background}

\subsection{Coding Agents}
\label{sec:landscape}

Coding agents are becoming a broader software engineering substrate rather than a narrow code-generation tool. After SWE-Bench~\citep{jimenez2024swebench}, issue-resolution work diversified into iterative single-agent systems~\citep{yang2024sweagent,zhang2024autocoderoverautonomousprogramimprovement,phan2024hyperagent}, multi-agent decompositions~\citep{tao2024magisllmbasedmultiagent,chen2024coderissueresolvingmultiagent}, structured workflows~\citep{xia2024agentless}, open platforms~\citep{openhands}, and bug-fix data synthesis for real-world resolution~\citep{pham2025swe}, as summarized by recent surveys~\citep{issue_resolution_survey,sase_survey}. Another line organizes agents around specifications or role-based collaboration rather than free-form prompts, as in Spec Kit~\citep{github2025speckit}, OpenSpec~\citep{fission2025openspec}, BMAD roles and story files~\citep{blackington2025bmad}, AgileCoder~\citep{nguyen2025agilecoder}, and Agentic Software Engineering with mentorship as code and lifecycle management~\citep{sase_survey}. At deployment scale, \citet{li2025aidev} analyse 456{,}000 pull requests from Codex, Devin, Copilot, Cursor, and Claude Code across 61{,}000 repositories and 47{,}000 developers, finding faster submission but lower acceptance than human pull requests; the proactivity gap sits inside this real-world acceptance gap. These lines now feed directly into developer-facing tools. Claude Code~\citep{anthropic2025claudecode}, Cursor~\citep{cursor2025}, Devin~\citep{cognition2024devin}, OpenAI Codex~\citep{openai2025codex}, Jules~\citep{google2025jules}, Gemini CLI~\citep{google2025geminicli}, Goose~\citep{block2025goose}, OpenCode~\citep{opencode2025}, Crush~\citep{charmbracelet2025crush}, Amazon Q Developer~\citep{amazon2024qdeveloper}, Cline~\citep{cline2025}, Aider~\citep{aider2024}, and OpenDev~\citep{bui2026building} place agents across developer workflows. They are the most relevant audit population here because they already support substantial autonomous work, yet public evidence remains thin on when they should initiate, interrupt, question, draft, or remain silent.

\subsection{Defining Proactivity}
\label{sec:defining_proactivity}

\citet{horvitz1999mixedinitiative} suggested that an interactive system should reason explicitly about three quantities: the expected benefit of acting for the user, the expected cost of interrupting the user, and the benefit of leaving the user in control. Following \citet{handler2023balancing} we use \emph{autonomy} for the property that the agent can act without supervision, and \emph{proactivity} for the property that the agent decides whether and when to act without an explicit prompt. \citet{agentsvsagentic2025} note that recent work often blends the two; we keep them separate throughout.

The \emph{suggestion} stream frames proactivity as anticipating useful tasks: ProactiveBench provides 6{,}790 human labeled events~\citep{luProactiveAgentShifting2024}; ContextAgent extends to multimodal sensory input~\citep{yangContextAgentContextAwareProactive2025}; ProAgent adds on demand sensing and reports a 33.4 point prediction gain~\citep{yangProAgentHarnessingOnDemand2025}; PROBE scores false interruptions and missed opportunities~\citep{pasternak2025probe}; and ProAgentBench uses 28{,}000 events from 500 hours of Microsoft 365 sessions with a When and How hierarchy~\citep{tang2026proagentbench}. The \emph{clarification} stream frames proactivity as asking before acting on uncertain intent: informative questions~\citep{gan2024clarqllm}, persistent user modeling for stateful SWE Bench and a 12 developer deployment~\citep{zhou2026tomswe}, joint productivity-proactivity-personalization training~\citep{sun2025trainingproactivepersonalizedllm}, deciding when hidden user preferences are relevant~\citep{kaur2026proper}, and a POMDP rule based on Expected Value of Perfect Information~\citep{suri2025sage}. Proactive coding should handle both streams across the development life cycle, and long-horizon work has begun to combine them through intent-conditioned monitoring, event-triggered follow-up, and deliberate \emph{stay silent} branches~\citep{shi2026chronosbench}.

\subsection{Reactive Substrates and Active User Simulation}
\label{sec:substrates}

The aspiration predates large language models: early agents targeted workload and information overload~\citep{maes1994agents}, direct manipulation concerns~\citep{shneiderman1997directmanipulation}, and continuously shown context~\citep{rhodes2000justintime,rhodes1996remembrance,rhodes1997wearableremembrance}. Current language model agents remain mostly reactive: ReAct loops~\citep{yao2023react}, agent computer interfaces~\citep{yang2024sweagentACI}, code action policies~\citep{wang2024codeact}, multi agent frameworks~\citep{wu2023autogen,park2023generativeagents}, and task benchmarks such as SWE-Bench~\citep{jimenez2024swebench}, $\tau$~\citep{yao2024taubench}, $\tau^2$~\citep{barres2025tau2bench}, and ARE~\citep{froger2025are} all assume that the developer formulates a task rather than asking whether it should have been raised.

The closest prior work to our protocol is PARE, which models applications as finite state machines, simulates users with constrained navigation, and gives user and agent different interfaces under a Stackelberg POMDP~\citep{nathani2026proactive,brero2024stackelbergpomdpreinforcementlearning}; PARE Bench covers 143 communication, productivity, scheduling, and lifestyle tasks, where frontier models top out near 42 percent.

\subsection{Interruption Cost in Software Development}
\label{sec:human_coding}

Poorly timed notifications can be net negative~\citep{mehrotra2016myphoneandme,okoshi2015reducingusers}. For developers, on-screen interruptions degrade code comprehension, with recovery ranging from 10--15 minutes for bug fixes to 30--60 minutes for architecture and security tasks~\citep{meyer2024breakingflow}; longitudinal evidence across 4{,}910 tasks and 17 developers further shows that self-interruptions can be more disruptive than external ones~\citep{abad2018taskinterruption}. Relevance Theory~\citep{sperber1995relevance}, Grice's Maxims~\citep{grice1975logic}, Active Inference~\citep{friston2010free}, and user intent or mental modeling methods~\citep{arora2024intentdetection,berkovitch2025identifyinggoals,lin2022inferringrewards,zhou2026tomswe} supply ingredients for relevance computation.

A state-aware IDE assistant reached 90 percent preference versus 47 percent for a persistent variant in a 65 participant study~\citep{chen2025need}. Feedback-delayed LLM code suggestions lifted acceptance from 4.9 to 18.6 percent and cut wasted inference calls by 75 percent over two months~\citep{alawad2025fdt}. Mental-state inference also helps code understanding~\citep{richards2024tomcode}, while CodingGenie implements a Level 2 fixed refresh policy and calls for further study~\citep{zhao2025codinggenie}. \citet{mozannar2024when} ground both lines in when to show a suggestion. Their taxonomies classify topics; our action space $\mathcal{A}$ classifies intents.

Both streams often assume the agent is invoked for a task and then exits. Proactive coding instead needs a sustained partner that runs across the development life cycle, carries memory across sessions, integrates with the team toolchain, and earns trust over time. Sustained presence enables situation-aware judgment, but not constant interruption. \citet{brady2026springdrift} report a 23 day deployment with append-only memory and ambient self-perception across email and web channels; it illustrates the always-on, not-always-speaking idea in a working system.

\section{A Three Level Taxonomy of Proactivity}
\label{sec: method}

We draw a three level taxonomy from mixed initiative theory~\citep{horvitz1999mixedinitiative} and autonomy taxonomies~\citep{handler2023balancing}. It gives shared vocabulary for comparing agentic coding systems and evaluations of proactive agents.

\subsection{A Decision Theoretic Formulation}
\label{sec: prob_form}

Let $s_t$ denote developer state at time $t$ (open buffer, branch, recent commits, calendar, sprint deadlines, ticket status, communication context), and let $\mathcal{E}_t$ denote the cross context event stream since the last decision. We group events into \emph{code}, \emph{project}, \emph{communication}, \emph{infrastructure}, and \emph{developer behavior}. Let $\mathcal{A}=\{\emph{notify},\emph{question},\emph{draft},\emph{stay silent}\}$ be the insight action space. The first three actions show an insight; \emph{stay silent} records a deliberate choice not to interrupt. The mixed initiative principles of \citet{horvitz1999mixedinitiative} can be read as:
\begin{equation}
a^\ast = \argmax_{a \in \mathcal{A}}\; \mathbb{E}_{p(o\mid a, s_t, \mathcal{E}_t)}\!\left[U(o; \theta)\right] \;-\; \mathrm{Cost}_{\text{int}}(s_t, a; \theta),
\label{eq:level3-rule}
\end{equation}
where $o$ is the developer-observable outcome of action $a$ (response, downstream code change, recovered context), $U(o; \theta)$ scores how useful that outcome is to the developer, with $\theta$ capturing what the agent has learned about this particular developer, and $\mathrm{Cost}_{\text{int}}(s_t, a; \theta)$ is the cost of breaking flow. The agent should \textbf{pick the action whose expected payoff to the developer beats the cost of interrupting them right now}, and pick \emph{stay silent} when no action clears that bar. Interruption cost varies by developer and workflow phase~\citep{meyer2024breakingflow,mehrotra2016myphoneandme,okoshi2015reducingusers}; coding examples include phase-aware gating from typing and error signals~\citep{chen2025need} and feedback-updated delay policies for LLM code suggestions~\citep{alawad2025fdt}. A practical $\mathrm{Cost}_{\text{int}}$ proxy could combine IDE focus state, idle time, edit cadence, test or incident activity, calendar status, and recent dismiss or defer signals, with local scoping for sensitive telemetry. Equation~\ref{eq:level3-rule} treats \emph{stay silent} as an explicit option, allowing one rule to cover suggestion and clarification.

\subsection{Three Levels}

We characterize an agentic coding system by which terms of Equation~\ref{eq:level3-rule} it actually takes into account. Figure~\ref{fig:levels} shows the levels, mapped into evaluation criteria O1 to O3 in Section~\ref{sec: experiments}.

\begin{itemize}[leftmargin=*,nosep]
\item \textbf{Level 1, Reactive.} The agent runs only when the developer starts an interaction. Between requests it has no persistent environmental presence, time model, or candidate action.
\item \textbf{Level 2, Scheduled.} The agent runs on a schedule or predefined event. It may filter, batch, rank, or summarize within that trigger, but does not learn a cross context, per developer interruption policy; silence is not a learned choice over the full event stream.
\item \textbf{Level 3, Situation-Aware.} The agent monitors $\mathcal{E}_t$ continuously, computes utility and interruption cost over the full action space, including \emph{stay silent}, and updates $\theta$ from developer responses. It decides what to show, when to show it, and whether showing anything is the right action.
\end{itemize}

\subsection{Insights as the Unit of Agent Output}
\label{sec:insights}

A Level 3 engine continuously evaluates Equation~\ref{eq:level3-rule} but does not always emit a message. Its action space is: \textbf{notify} (state change), \textbf{question} (clarification), \textbf{draft} (pull request comment, patch, or review thread), and \textbf{stay silent} (a deliberate choice not to interrupt). An insight is a context grounded, time sensitive hypothesis about what matters next, paired with one action. An \emph{insight policy} selects the action, chooses evidence, frames any message, and updates future decisions from feedback. The metrics in Section~\ref{sec: experiments} score this stream rather than task completion. The suggestion stream~\citep{luProactiveAgentShifting2024,yangContextAgentContextAwareProactive2025,yangProAgentHarnessingOnDemand2025} mainly measures \emph{notify} and \emph{draft}; the clarification stream~\citep{gan2024clarqllm,sun2025trainingproactivepersonalizedllm,zhou2026tomswe} mainly measures \emph{question} and \emph{stay silent}. A proactive coding agent, in the spirit of \citet{horvitz1999mixedinitiative}, should handle all four.

\stitle{Acceptance criteria} A candidate insight should satisfy four checks. First, it should be relevant now: the expected developer benefit should exceed interruption cost, and the policy should choose silence when the same fact can safely wait. Second, it should be grounded: a reviewer should be able to recover the files, diffs, tickets, logs, messages, or behavioral signals that support the claim. Third, it should be action matched: \emph{notify} for awareness, \emph{question} for missing intent, \emph{draft} for low ambiguity work, and \emph{stay silent} for low value or poorly timed items. Fourth, it should be learnable: acceptance, dismissal, deferral, edits, and later delegation should change future timing or framing. These checks turn proactivity from a product label into behavior that can be labeled and tested.

The Level 3 claim is meant to be tested. It would be revised by evidence that a coding agent computes a meaningful $\mathrm{Cost}_{\text{int}}$ from developer state, uses \emph{stay silent} as an explicit action, and updates $\theta$ from per developer feedback (O1 - O3, Section~\ref{sec: experiments}).

\section{Comparing the Landscape and Sketching an Evaluation Protocol}
\label{sec: experiments}

Section~\ref{sec:audit} compares five agentic coding systems against five practical criteria from Section~\ref{sec: method}. Section~\ref{sec:pare-coding} adapts active user simulation~\citep{nathani2026proactive} to software development, and Section~\ref{sec:metric} proposes scoring the resulting insight stream rather than task completion. We present no measured results.

\subsection{Side by Side Comparison of Five Contemporary Coding Agents}
\label{sec:audit}

We compare five widely deployed agentic coding systems and one reference architecture in Table~\ref{tab:audit} using verifiable claims from technical disclosures or reputable reporting: cloud agents, issue-to-pull-request agents, scheduled and suggested tasks, webhook and integration triggers, memory tools, SWE Bench Verified performance, and ambient-agent reference architectures~\citep{cursor_automations2026,jules2025,google_jules2025,jules_changelog,techcrunch_cursor2026,anthropic_routines2026,anthropic2026trends,langchain_ambient2025,langchain_agents_from_scratch2025}. We restrict this to software-development systems, while treating adjacent systems with feedback driven topic policy or desktop scheduling as comparators~\citep{openai_pulse2025,claude_cowork2026}.

We use the following, marking \cmark{} for clear documentation, \pmark{} for partial evidence, and \xmark{} when no description is available. O1 to O3 are the testable conditions from Section~\ref{sec:insights}; O4 and O5 are practical requirements for comparison.
\begin{itemize}[leftmargin=*,nosep]
\item \textbf{O1, Cost of interruption is computed}. The system observes developer state and decides when to show a message.
\item \textbf{O2, Stay silent is an explicit action} (the inclusion of \emph{stay silent} in the action space $\mathcal{A}$). The system can detect a candidate action and still choose not to show it.
\item \textbf{O3, Per developer feedback updates the policy}. The system records developer responses and changes future decisions about what to show and when.
\item \textbf{O4, Cross context observation}. The system ingests events from at least three categories of context (Section~\ref{sec: prob_form}).
\item \textbf{O5, Initiation channel}. The system can show messages without requiring the developer to open a separate tool.
\end{itemize}

\begin{table}[H]
\centering
\caption{Comparison of five deployed coding agents and one reference architecture. \cmark{}: clearly documented; \pmark{}: partial; \xmark{}: not found. $^\dagger$Conceptual reference architecture.}
\label{tab:audit}
\small
\setlength{\tabcolsep}{5pt}
\begin{tabular}{l c c c c c c}
\toprule
\textbf{System} & \textbf{Level} & \textbf{O1} & \textbf{O2} & \textbf{O3} & \textbf{O4} & \textbf{O5} \\
\midrule
Cursor Background Agents~\citep{cursor_automations2026} & 1 & \xmark & \xmark & \xmark & \pmark & \cmark \\
GitHub Copilot Coding Agent & 2 & \xmark & \xmark & \xmark & \pmark & \cmark \\
Jules~\citep{jules2025,google_jules2025} & 2 & \xmark & \xmark & \xmark & \pmark & \cmark \\
Cursor Automations~\citep{cursor_automations2026,techcrunch_cursor2026} & 2 & \xmark & \xmark & \pmark & \cmark & \cmark \\
Claude Code Routines~\citep{anthropic_routines2026} & 2 & \xmark & \xmark & \xmark & \pmark & \cmark \\
\midrule
LangChain Ambient Agents~\citep{langchain_ambient2025} & 3$^\dagger$ & \pmark & \cmark & \cmark & \pmark & \cmark \\
\bottomrule
\end{tabular}
\end{table}

To the best of our finding in surveyed public materials, no deployed coding agent documents a meaningful interruption cost or explicit silence action. O3 varies only weakly: Cursor Automations exposes memory, but its effect on what is surfaced is unclear. Adjacent products show feedback driven and presence gated variants of O1 and O3~\citep{openai_pulse2025,claude_cowork2026,zheng2025persono}, so the gap lies in adapting existing interaction patterns to the specificities of software development, not inventing entirely new or non-existent capabilities. O5 is widely satisfied; the frontier is to trigger breadth, not show judgment.

\subsection{Adapting Active User Simulation to Software Development Workflows}
\label{sec:pare-coding}

\citet{nathani2026proactive} observe that proactivity cannot be measured on static traces because timing depends on user behavior. Their PARE framework simulates the user under a Stackelberg POMDP~\citep{brero2024stackelbergpomdpreinforcementlearning}. A coding adaptation would expose the event categories, like the ones in Section~\ref{sec: prob_form}, as state changes and condition a simulated developer on role, commitments, interruption tolerance, and workload.

ClawBench~\citep{zhang2026clawbench} reinforces the need for realistic, multi-step evaluation with traceable trajectories. A coding focused long-horizon benchmark should keep rich event traces but score insight decisions rather than final task completion.

\stitle{Long-horizon structure} In this setting, long-horizon should not mean that the agent simply runs for many steps. It should mean that useful evidence arrives over time and the right action can change as the situation develops. A scenario should include early weak signals, distractors that should be ignored, later confirming evidence, a moment where silence is correct, and a later response that should affect future behavior. This structure tests whether an agent can wait when evidence is thin, connect signals across tools when the case becomes stronger, and remember feedback without overreacting to one dismissal or acceptance.

In such a benchmark, each scenario would define states $\{s_t,\mathcal{E}_t\}_{t=1}^T$, a reference action $a_t^\star \in \mathcal{A}$, support facts $G_t^\star$, and feedback $f_t$ for shown insights. The agent returns $a_t$ and, for \emph{notify}, \emph{question}, or \emph{draft}, a message $m_t$ with recoverable grounding $G_t$. A seed suite across API breakage, dependency lifecycle, incident response, and routine background work should include silence, showing a message, and later feedback use.

\stitle{Trace collection} Data should be collected as timestamped, replayable developer-workflow traces rather than as completed task records. Each trace should align repository diffs, issue and pull request events, CI logs, dependency or security notices, communication snippets, and privacy-scoped IDE state such as idle time, active file, edit cadence, and test cadence into a single event ledger. Benchmark creators would then sample decision points from both candidate alerts and quiet intervals, because the denominator for proactivity includes opportunities where the correct action is to remain silent. These traces can be synthesized from public repositories for seed scenarios, but deployment studies should log only scoped metadata or redacted excerpts needed to judge timing and grounding.

\stitle{Example scenario} A payment provider announces that an older API version will be retired in two weeks. At first, the reference action is \emph{stay silent}: the warning appears only in a release note, the repository has no failing tests, and the developer is working on an unrelated incident. Later, the agent observes that the developer's current branch touches the checkout flow, a Jira ticket in the sprint depends on the same endpoint, CI starts showing deprecation warnings, and a Slack thread confirms that the team plans to migrate this week. The reference action becomes \emph{notify}, grounded in the release note, affected call sites, Jira ticket, CI warning, and Slack thread. If the developer asks the agent to handle routine migration work, a later state can make \emph{draft} the correct action. If the developer dismisses low-priority dependency alerts during focus blocks, LL tests whether the agent learns to delay similar messages while still showing urgent migration risks.

\subsection{Evaluating the Insight Stream}
\label{sec:metric}

SWE Bench~\citep{jimenez2024swebench} scores what an agent does once a task is given. A Level 3 engine instead produces an insight stream that may be acted on, ignored, or never shown. We score whether that stream keeps the agent on the right path. Let $t$ index a decision point, $T$ the number of decision points in a scenario, $s_t$ the current developer and project state, $\pi$ the surfacing policy being evaluated, $\mathcal{A}=\{\emph{notify}, \emph{question}, \emph{draft}, \emph{stay silent}\}$ the action space, $a_t^\pi \in \mathcal{A}$ the action selected by that policy, and $a_t^\star$ the reference action.
\begin{itemize}[leftmargin=*,nosep]
\item \textbf{Insight Decision Quality (IDQ)} scores whether the agent chose the right action at the right time:
\begin{equation}
\mathrm{IDQ}(\pi)=\frac{1}{T}\sum_{t=1}^{T} S_{\mathrm{dec}}(a_t^\pi,a_t^\star,s_t).
\end{equation}
Here $S_{\mathrm{dec}}(a_t^\pi,a_t^\star,s_t)\in[0,1]$ is the decision score at time $t$. It gives full credit to the reference action, partial credit to reasonable but weaker alternatives, and penalties for false interruptions, missed opportunities, wrong action types, and bad timing. When the policy is clear from context, we write $\mathrm{IDQ}$ for $\mathrm{IDQ}(\pi)$. Unlike acceptance rate, IDQ scores both shown insights and deliberate silence.
\item \textbf{Context Grounding Score (CGS)} asks whether a shown insight is supported by the right evidence:
\begin{equation}
\mathrm{CGS}=\frac{1}{|\mathcal{M}|}\sum_{t\in\mathcal{M}} F_1(G_t,G_t^\star)\cdot \mathds{1}[\mathrm{faithful}(m_t,G_t)],
\end{equation}
where $\mathcal{M}$ is the set of times for which $a_t^\pi$ is not the \emph{stay silent} action, $|\mathcal{M}|$ is the number of shown insights, $G_t$ and $G_t^\star$ are the agent and reference support facts, $F_1(G_t,G_t^\star)$ is the harmonic mean of evidence precision and recall, and $\mathds{1}[\mathrm{faithful}(m_t,G_t)]$ equals 1 only when the factual claims in $m_t$ are supported by $G_t$. If $\mathcal{M}$ is empty, CGS should be reported as not applicable rather than forced to zero.
\item \textbf{Learning Lift (LL)} measures whether feedback improves later decisions:
\begin{equation}
\mathrm{LL}=\mathrm{IDQ}(\pi_{\mathrm{adapted}})-\mathrm{IDQ}(\pi_{\mathrm{frozen}}).
\end{equation}
Here $\pi_{\mathrm{adapted}}$ may update from earlier developer feedback $f_t$, while $\pi_{\mathrm{frozen}}$ cannot. Both are evaluated on the same later decision points, so positive LL means feedback improved future timing or action choice rather than merely repeating the previous message.
\end{itemize}
Together, the metrics ask whether the agent chose the right action, used the right evidence, and improved after feedback. Repository outcomes remain useful secondary checks, but the primary unit should remain the insight decision.

\stitle{Annotation protocol} A useful benchmark requires two labeling layers. \emph{Action selection} labels the reference action $a_t^\star$ and acceptable alternatives at each decision point. \emph{Grounding verification} labels support facts $G_t^\star$ and factual claims in $m_t$, making CGS checkable when an insight omits critical context or uses an alternate but sufficient evidence set. Both layers should allow multiple correct answers because developers differ in interruption tolerance and preference for detail.

\stitle{Metric validity} IDQ, CGS, and LL should be validated as repeatable labeling protocols. LLM labelers can pre-segment traces, propose actions and support facts, and flag disagreements, but human reviewers should audit final labels and report agreement. Traces should include recurring scenarios so LL can compare adapted and frozen policies on held out later decisions.

\stitle{Policy interpretation} Scores should be read together. High IDQ and low CGS means the agent has timing judgment but weak justification; high CGS and low IDQ means it can ground messages but surfaces them poorly. Positive LL shows useful adaptation, while negative LL reveals overfitting to feedback.

\stitle{Validity limits} These metrics are evaluation targets rather than a completed benchmark. A full benchmark would need reference insight labels for IDQ, support facts for CGS, recurring but non-identical situations for LL, annotator agreement, and held out variants for measuring learning. The metrics should be reported together rather than collapsed into one ranking number.

\section{Discussion}
\label{sec: analysis}

The framework above makes proactivity testable before a fully deployed Level 3 coding agent exists. The key question is not only whether an agent completes more tasks, but whether it notices the right situation, chooses an appropriate moment, explains its evidence, and learns from feedback. This shifts attention from task completion alone to the infrastructure needed to observe candidate insights, silent decisions, and later developer responses.

\stitle{Timing and silence} Interruption cost is not a constant property of a notification; it depends on what the developer is doing, how urgent the event is, and how expensive recovery would be~\citep{meyer2024breakingflow}. A proactive agent therefore needs a learned $\mathrm{Cost}_{\text{int}}$ estimate that uses observable signals across tools and avoids cold-start messages whose value is lower than their attention cost. A practical study could compare three policies: show every detected issue, show only after an idle threshold, and show only when learned $\mathrm{Cost}_{\text{int}}$ permits it. The important design choice is to log both shown messages and candidate messages that were withheld, since silence is meaningful only if the evaluator can see what the agent declined to surface. IDQ, CGS, and LL would then measure timing, grounding, and adaptation.

\stitle{Communication} A shown insight also has to be framed at the right level of detail. Code-review automation suggests that comment quality matters more than volume, but proactive agents still need a measurable account of when evidence helps and when it increases reading burden. A benchmark could compare a headline, a short reason, a diff-grounded reason, and an interactive version that lets the developer request more context. IDQ, CGS, and time to decision would expose both under-explaining and over-explaining.

\stitle{Benchmarking} Existing benchmarks provide important pieces, but none evaluates monitoring, relevance, timing, framing, and silence as one combined behavior. ProactiveBench~\citep{luProactiveAgentShifting2024}, ContextAgent~\citep{yangContextAgentContextAwareProactive2025}, and ProAgent~\citep{yangProAgentHarnessingOnDemand2025} emphasize suggestion; \citet{gan2024clarqllm} and \citet{zhou2026tomswe} emphasize clarification; PARE Bench~\citep{nathani2026proactive} studies general productivity; and ARE~\citep{froger2025are} and $\tau^2$~\citep{barres2025tau2bench} support reactive-agent evaluation. A coding benchmark should release replayable traces, IDQ, CGS, and LL labels, and comparable implementations where public APIs permit. The trace format matters: repository, issue, CI, communication, and privacy-scoped IDE signals should be aligned by time so that evaluators can judge what the agent knew before it acted. The result should support diagnosis rather than reduce developer experience to one leaderboard number.

\stitle{Evaluation data collection} The most useful evaluation data would come from prospective logging rather than retrospective issue mining. During normal work, the system should record a bounded event ledger, run the insight policy in shadow mode, and store candidate decisions even when nothing is shown. For each sampled point, annotators should see only the evidence available before that time, then label the best action, acceptable alternatives, required support facts, and whether a developer response later changed the preferred policy. This design avoids two common shortcuts: treating every completed task as a missed proactive opportunity, and treating every alert dismissal as proof that the alert was wrong. It also gives LL a real denominator, because the benchmark can compare what the adapted policy would do after feedback with what a frozen policy would have done on related later events. For privacy, raw IDE snapshots and chat text need not be released; benchmarks can store hashed event identifiers, redacted evidence windows, and provenance for each visible excerpt. Evaluation logging can be broader than what the product surfaces, but labels must preserve temporal fidelity and use only information available before the decision.

\stitle{Product design} A Level 3 coding agent should be an accountable decision surface, not merely a more persistent Level 2 automation. Its interface needs an inbox or agent-management view that makes shown and deferred insights inspectable, lets developers accept, edit, defer, or dismiss suggestions, and records those responses as feedback. The interface should also show enough evidence for the developer to understand why an insight appeared now, while keeping detailed behavioral telemetry out of team-visible channels. This requires a boundary between private developer state and shared project context: interruption tolerance and dismissal patterns should remain private, while evidence such as failing tests, affected files, and linked issues should remain auditable.

\stitle{Reporting requirements} Future systems should report more than task success and trigger coverage. A credible proactivity claim should state which event streams were observed, how candidate insights were filtered, how often the system chose to \emph{stay silent}, what feedback was collected, and whether feedback changed later surfacing decisions. Silence rates, delayed-message rates, feedback categories, and learning lift are most useful with the denominator of candidate insights considered but not shown.

\stitle{Limitations} This paper is a position and measurement proposal. The audit relies on public documentation, and O1 to O5 are practical criteria rather than formal guarantees. Simulated developers can diverge from real users~\citep{sun2025trainingproactivepersonalizedllm,zhou2026tomswe}, and interruption findings from real-time work may not transfer cleanly to asynchronous agent messages~\citep{meyer2024breakingflow}. These limits are reasons to report uncertainty and treat proactivity as a property involving real developers rather than a benchmark score alone.

\section{Conclusion}
\label{sec: conclusion}

Coding agents are maturing from tools that execute prompted tasks into partners that may notice drift, uncertainty, or emerging opportunities before being asked. Proactivity in agentic coding should be judged not by how often an agent acts, but by whether it surfaces the right insight at the right time, with enough evidence, and stays silent when intervention is unwarranted. Our three level taxonomy separates reactive execution, scheduled triggering, and situation-aware judgment; the Level 3 engine sketch centers the insight as the unit of proactive behavior; and IDQ, CGS, and LL make that behavior measurable. Together, these pieces frame proactive coding as accountable mixed initiative collaboration grounded in timing, relevance, evidence, and trust.

\bibliographystyle{abbrvnat}
\ifuseneurips
\else
\nobibliography*
\bibliography{custom}
\fi

\end{document}